\documentclass[aps,prd,twocolumn,superscriptaddress]{revtex4}
\usepackage{graphicx}
\usepackage{amsmath}
\usepackage{amssymb}
\usepackage{slashed}
\usepackage{latexsym}
\usepackage{epsfig}
\usepackage{amsbsy}
\usepackage{array}
\usepackage{amssymb}
\usepackage{color}
\usepackage{bm}
\usepackage{lipsum}
\usepackage{mathrsfs}
\usepackage{float}
\usepackage{color}

\begin{document}

\title{ Two-flavor color superconductivity in the Fierz-transfromed Nambu$ - $Jona-Lasinio model}

\author{Wen-Hua Cai}
 \email{stuwenhua.cai@gmail.com }
 \affiliation{College of physics, Sichuan University, Chengdu 610064, China}

\author{Qing-Wu Wang }
 \email{qw.wang@scu.edu.cn }
 \affiliation{College of physics, Sichuan University, Chengdu 610064, China}

\begin{abstract}
 The color superconducting phase of two-flavor quark matter is studied  under a Fierz-transformed Lagrangian.  In the Fierz-transformed Lagrangian both the quark-antiquark and quark-quark channel are included. Two parameters  $\alpha$ and $\beta$ are introduced to weight the Fierz-transformed quark-antiquark Lagrangian and  quark-quark Lagrangian, respectively. The  couplings of different interaction channel   are thus fixed by the two coefficients other than five free parameters.  The interplay between chiral symmetry restoration and the formation of color superconducting phases are discussed.  It is found that the increase of $\alpha$ leads the chiral phase transition from first order to crossover. The calculated specific heat   shows jump at critical temperature which confirms the phase transition.
The increase of $\beta$  leads the pseudo-critical chemical potential reduce. With large $\beta$ the superconductors gap $\Delta$ can be larger than the typical value and the well-known BCS relation between critical temperature and gap is well fitted.

\end{abstract}

\maketitle
\section{Introduction}

Investigation of deconfined quark matter is a hot issue in the study of the strong interaction. The quark-gluon plasma (QGP) which  can be reproduced by relativistic heavy ion collisions (RHICs) experiment \cite{Adams:2005dq, Shuryak:2008eq}  may exists as  gluons deconfine in the adjoint representation at high temperature.
At low temperature and less baryon density, hadrons and mesons are confined that all states are color singlets.
 The phase transition from hadronic matter to the QGP is a broad crossover at zero baryon chemical potential. Although the transition property at high densities is still debated, but it is widely expected to be first-order \cite{Nahrgang:2016ayr, Bzdak:2019pkr, Dexheimer:2020zzs}  and evidences in study of the EOS of massive compact stars show that quark-matter cores may exist in the stars and it could be a first order phase transition \cite{Annala:2019puf, Bauswein:2018bma}.

 The compact star mass and radius heavily depended on the quark core mass and radius.  But the regime of low
temperatures and moderate densities is not accessible by perturbative QCD up to now, and  the QCD critical point is uncertain.   There are so many uncertainties in connecting the hadron phase and quark phases.  In the schematic phase diagram,  there is a possible phase transition between the nuclear superfluid and color superconductors \cite{Fukushima:2010bq, Fukushima:2013rx} at sufficiently large densities and low temperatures. Many kinds of pairing patterns and condensate exist, such as meson, diquark and four-quark condensate \cite{Bailin:1983bm, Evans:1999at,Rajagopal:2000ff, Alford:2007xm,Alford:2007qa}.
In-depth research on hadron    and quark color super-conductance and super-fluidity,  which provide useful information on the phase transition \cite{Alford:2001zr,  Buballa:2003qv} and  viscous and diffusive effects near the QCD critical point \cite{Monnai:2016kud,Monnai:2017ber,Fotakis2020prd,Alford:2008pb}, could have potential applications to heavy  ion collisions and  astrophysics of neutron stars   \cite{Pisarski:1999tv,Popov:2005xa, Ruester:2003zh,Roupas:2020nua,Cai:2020xnz}.

For the quark matter, quark condensate plays a key role in quark mass formation  and phase transitions. With the spontaneous chiral symmetry breaking the quark-antiquark condensates provide the main contribution of the effective mass.  In analogy with Cooper pairs of electrons with opposite momentum and opposite spin, cold deconfined quarks could become    paired to form  color superconductor  with diquark condensate. To study color superconductor it can be started by applying the color current interaction $-g(\bar \psi \gamma_{\mu}\lambda_a\psi)^2$ directly, with a free coupling $g$ which is the common practice.  After taking the mean field approximation, only the scalar current  $-g(\bar \psi \gamma_{0}\lambda_a\psi)^2$ remains. To incorporate into more interaction channel it always resorts to the  Fierz transformation of the original colore current interaction \cite{Klevansky:1992qe, Buballa:2003qv, Su:2020xka}.
On the opposite side,
 the standard NJL Lagrangian   contains only scalar and pseudo-scalar interactions  but without the colore current interaction. However, with the Fierz transformation we can produce color current interaction and the vector interaction with which the chemical potential needs to be revised.

 The Fierz transformation is a mathematically equal description of the original Lagrangian which means the original Lagrangian and its Fierz transformation should be all contained in the complete Lagrangian with appropriate weight. Besides the color current interaction,   
 with the rearrangement of the position of quark and antiquark, the Fierz transformed  Lagrangian contains an additional scalar term which  will modify the contribution of the scalar four-point interaction in the standard Lagrangian. 
 
 The introduce of weighting parameters of the original Lagrangian and its Fierz transformations is not equal to redefine the couplings of variant interaction channel.
 In our previous paper, the introduction of weighting parameter seems   mathematically equal to a redefinition of the coupling of vector channel interaction \cite{Wang:2019uwl,Li:2018ltg, Wang:2019jze, Wang:2021mfj}. In Ref. \cite{Wang:2019jze},
 the modification to the effective chemical potential is written  as $\mu_{r1}=\mu-\alpha n$  $G^{\prime}/[N_c(1-\alpha)+\alpha]$  where $\alpha$ is the weighing parameter,  $G^{\prime}$ is determined by the low energy experiment data and  $n$ is the vector condensate.
 If the  original Lagrangian has  a vector channel interaction in the form of $-G_V(\bar q \gamma_{\mu} q)^2$, the effective chemical potential is $ \mu_{r2}=\mu-2G_V n$.
As compared the expressions of $\mu_{r1}$ and $\mu_{r2}$,  it is easy to see that  the adjusting of weighting parameters $\alpha$ is equal to adjust the vector coupling $G_V$. But  this kind of equivalence is a result  that we have   neglected the color interaction terms $G(\bar q \gamma_0\lambda_0 q)^2$ in the Fierz-transformed Lagrangian for a color neutral system.  Also we will show latter that   the Fierz transformation gives a fixed ratio of couplings for some interaction channels which is different from any other works.

In this paper, we study the quark-antiquark and diquark condensate from a  Lagrangian that contains standard Lagrangian and its Fierz transformation with quark-quark and quark-antiquark interactions   to investigate their competition and mutual influence.
To keep a global antisymmetry of the wave function, the diquark condensates can  involve  two colored quarks with spin-0  or with spin-1 for a single color. The estimated gap  of the vector channel is much smaller  than the scalar channel \cite{Alford:1997zt}.
In this paper we focus on  the  two-flavor superconductors (2SC) quark matter with spin-0.  The paper is organized as follows. In Sec. \ref{sec.mod}, we present the NJL model. We derive the Fierz-transformed Lagrangian and the solutions in the Nambu-Gorkov formalism. In Sec. \ref{sec.result}, we show our result calculated at different weight of the vector channel and diquark channel.  And in the end we give a short summary.

\section{Model } \label{sec.mod}

The standard NJL Lagrangian with four point interaction is

\begin{equation}
\mathscr{L}=\bar{\psi}(\mathrm{i}\not \partial-m) \psi+g\left\{(\bar{\psi} \psi)^{2}+\left(\bar{\psi} \mathrm{i} \gamma_{5} \vec{\tau} \psi\right)^{2}\right\}.
\end{equation}
Under the mean field approximation, only the scalar channel remains which governs the chiral restoration.  When considering the contribution of the vector channel, the usual method is to introduce a vector  coupling constant. Similarly,  the conventional way to study the color superconductor phase  is to consider color current interaction with introduction of color current coupling or the color electric and magnetic coupling constants. But all these terms  exist in the  Fierz transformation of the standard NJL Lagrangian.
Taking into account the anti-commutation rules for fermions, the four Fermi interaction

\begin{equation}
\mathscr{L}_{int}=g_I(\bar q \hat \Gamma q)^2=g_i\Gamma_{i j}\Gamma_{k l}\bar{ q_i}q_j\bar{ q_k}q_l
\end{equation}
change to
\begin{eqnarray}
\mathscr{L}_{ex} =-g_i\Gamma_{i l}\Gamma_{k j}\bar{ q_i}q_l\bar{ q_k}q_j
\end{eqnarray}
and
\begin{equation}
\mathscr{L}_{qq} =g_i\Gamma_{i j}\Gamma_{k l}\bar{ q_i}\bar{ q_k}q_lq_j.
\end{equation}
The three interactions, $\mathscr{L}_{int}$, $\mathscr{L}_{ex}$ and  $\mathscr{L}_{qq}$,  are mathematically equal, but apparently   the contributions of different interaction channels vary under the mean field approximation.

In order to make the Lagrangian under the mean field approximation as close as possible to the original interaction, we take the following Lagrangian
\begin{equation}\label{lcoup}
\mathscr{L}=\bar{\psi}(\mathrm{i}\not \partial-m) \psi+ (1-\alpha-\beta)\mathscr{L}_{int}+\alpha \mathscr{L}_{ex}+\beta\mathscr{L}_{qq}.
\end{equation}
  In this way, the  vector channel and color current interactions are naturally included. Although the interactions are still included  arbitrary numbers, the meanings are different.  The $\alpha$ and $\beta$   weight the contribution from different interactions and set up constraints  on the vector channel and diquark channel. Different from models that staring from the  color current interactions, the   coupling ratio  $G_S/H_S$ of quark-antiquark and diquark is no longer a constant \cite{Buballa:2003qv}.

\subsection{The Fierz transformations}

The Fierz tansformation is under  spinor, flavor and color spaces.
The exchange diagrams lead to  $  \mathscr{L}_{ex} $  for the $ \bar q q$ interactions and for the scalar and pseudo-scalar channel we have
operators of the interactions
\begin{equation}
\begin{array}{l}
{[s]_{\alpha \beta^{\prime} ; \alpha^{\prime} \beta} = \frac{1}{4}\left[s-p+v-a+\frac{1}{2} t\right]_{\alpha \beta ; \alpha^{\prime} \beta^{\prime}}} \\
{[p]_{\alpha \beta^{\prime} ; \alpha^{\prime} \beta} = \frac{1}{4}\left[-s+p+v-a-\frac{1}{2} t \right]_{\alpha \beta ; \alpha^{\prime} \beta^{\prime}}}
\end{array}
\end{equation}
where
\begin{equation}
\begin{array}{l}
s_{\alpha \beta ; \alpha^{\prime} \beta^{\prime}}=1_{\alpha \beta} 1_{\alpha^{\prime} \beta^{\prime}} \\
p_{\alpha \beta ; \alpha^{\prime} \beta^{\prime}}=\left(i \gamma_{5}\right)_{\alpha \beta}\left(i \gamma_{5}\right)_{\alpha^{\prime} \beta^{\prime}} \\
v_{\alpha \beta ; \alpha^{\prime} \beta^{\prime}}=\left(\gamma_{\mu}\right)_{\alpha \beta}\left(\gamma^{\mu}\right)_{\alpha^{\prime} \beta^{\prime}} \\
a_{\alpha \beta ; \alpha^{\prime} \beta^{\prime}}=\left(\gamma_{\mu} \gamma_{5}\right)_{\alpha \beta}\left(\gamma^{\mu} \gamma^{5}\right)_{\alpha^{\prime} \beta^{\prime}} \\
t_{\alpha \beta ; \alpha^{\prime} \beta^{\prime}}=\left(\sigma^{\mu v}\right)_{\alpha \beta}\left(\sigma_{\mu \nu}\right)_{\alpha^{\prime} \beta^{\prime}}
\end{array}
\end{equation}

In the quark-antiquark channel, the transfromation matrix in the $U(N)$ symmetry is

\begin{equation}\left(\begin{array}{c}
(1)_{i j}(1)_{k l} \\
\left(\tau_{a}\right)_{i j}\left(\tau_{a}\right)_{k l}
\end{array}\right)=\left(\begin{array}{cc}
\frac{1}{N} & \frac{1}{2} \\
2 \frac{N^{2}-1}{N^{2}} & -\frac{1}{N}
\end{array}\right)\left(\begin{array}{c}
(1)_{i l}(1)_{k j} \\
\left(\tau_{a}\right)_{i l}\left(\tau_{a}\right)_{k j}
\end{array}\right),
\end{equation}
with $a=1,2,\cdots, N^2-1.$
So in the flavor space with $ N=2 $
\begin{eqnarray}
1\cdot 1&\to&\frac{1}{2}1\cdot 1+\frac{1}{2} \tau \cdot \tau,\\
\tau\cdot \tau&\to&\frac{3}{2}1\cdot 1-\frac{1}{2} \tau \cdot \tau.
\end{eqnarray}
Here all the subscripts are neglected for simplification.
And in the color space with  $N=3$
\begin{eqnarray}
\lambda_0\cdot \lambda_0&\to&\frac{1}{3}\lambda_0\cdot \lambda_0+\frac{1}{2} \lambda \cdot \lambda .
\end{eqnarray}
Here, only transfromation of the  color single was considered and more details can refer to the appendix of \cite{Buballa:2003qv}.
A full transformation under  spinor, flavor and color spaces results in
\begin{equation} \begin{array}{l}
~~~~~~\frac{1}{4}[s-p+v-a+\frac{1}{2} t]\otimes(\frac{1}{2}1\cdot 1
+\frac{1}{2} \tau \cdot \tau)  \otimes (\frac{1}{3}\lambda_0\cdot \lambda_0
 \\~~~~ +\frac{1}{2} \lambda \cdot \lambda)
    +\frac{1}{4}[-s+p+v-a-\frac{1}{2} t]
     \otimes(\frac{3}{2}1\cdot 1 -\frac{1}{2} \tau \cdot \tau)\\~~~~\otimes(\frac{1}{3}\lambda_0\cdot \lambda_0+\frac{1}{2} \lambda \cdot \lambda )\\
=-[\frac{1}{8}(2s-2p-4v+4a+t)\otimes1\cdot 1\\
+(-2s+2p-t)\otimes\tau \cdot \tau]\otimes(\frac{1}{3}\lambda_0\cdot \lambda_0+\frac{1}{2} \lambda \cdot \lambda).
\end{array}
\end{equation}






In the quark-quark channel, the transfromation matrix in the $  U(N) $ symmetry is

\begin{equation}\left(\begin{array}{c}
(1)_{i j}(1)_{k l} \\
\left(\tau_{a}\right)_{i j}\left(\tau_{a}\right)_{k l}
\end{array}\right)=\left(\begin{array}{cc}
\frac{1}{2} & \frac{1}{2} \\
\frac{N -1}{N } & -\frac{N+1}{N}
\end{array}\right)\left(\begin{array}{c}
(\tau_{S})_{i k}(\tau_{S})_{l j} \\
\left(\tau_{A}\right)_{i k}\left(\tau_{A}\right)_{l j}
\end{array}\right).
\end{equation}
Here, the subscripts  $ S $ and $ A $ to show that the generators are symmetry or anti-symmetry under transposition.  The total transformation to quark-quark channel is

\begin{equation} \label{eq.color}
	\begin{array}{c}
 \frac{1}{4}[s-p+v-a-\frac{1}{2} t]\otimes(\frac{1}{2}1\cdot 1+\frac{1}{2} \tau \cdot \tau)\\
 \otimes(\frac{1}{3}\lambda_S\cdot \lambda_S+\frac{1}{2} \lambda_A \cdot \lambda_A)  + \frac{1}{4}[-s+p+v \\ -a+\frac{1}{2} t]
\otimes(\frac{3}{2}1\cdot 1-\frac{1}{2} \tau \cdot \tau\otimes)(\frac{1}{3}\lambda_S\cdot \lambda_S \\ +\frac{1}{2} \lambda_A \cdot \lambda_A).
\end{array}
\end{equation}
The most important diquark channel are the $  qi\gamma_5C\tau_2\lambda_A q$  and   $  q\gamma_0\gamma_5C\tau_2\lambda_A q$. But in Eq. \eqref{eq.color}, the colored vector channel is cancelled out. Then we have only the colored scalar channel which read as
 \begin{equation}
\frac{1}{2}s\otimes\frac{1}{2}\tau_2\cdot\tau_2\otimes \frac{1}{2}\lambda_A\cdot \lambda_A.
\end{equation}
Under the mean field approximation, only the scalar, vector and   diquark channel remain which are listed here with
\begin{eqnarray}
 \mathscr{L}_{int}&=&g(\bar  q q)^2,\\
\mathscr{L}_{ex}&=& \frac{g}{12}[(\bar  q q)^2-2(\bar  q\gamma_0 q)^2]+ \frac{g}{8}[(\bar  q\lambda_0 q)^2 \nonumber\\
&&\qquad-2(\bar  q\gamma_0\lambda_0 q)^2],\\
 \mathscr{L}_{qq}&=& \frac{g}{8}((\bar  qi\gamma_5C\tau_2\lambda_A\bar q)(   qi\gamma_5C\tau_2\lambda_A q)).
\end{eqnarray}
 Considering the weighting parameters in \eqref{lcoup},
 the effective couplings for different interaction channels are

\begin{eqnarray}
G_s^{(0)} =(1-\alpha-\beta)g+\frac{\alpha}{12}g,&& G_s^{(8)} =\frac{\alpha}{8}g,\\
G_v^{(0)}=-\frac{\alpha}{6}g,\       G_v^{(8)}=-\frac{\alpha}{4}g,&&\
H=  \frac{\beta}{8}g.
\end{eqnarray}
Here, the subscript $s$ is short for scalar and $v$ for vector; the superscript $0$ and $8$ is to distinguish between uncolored and colored channels.
The coupling $H$ is of quark-quark interaction.  It is apparent that the introduction of  weigh parameters $ \alpha $ and $ \beta $ can not be the same thing  with simply adjusting the couplings,  since we have five couplings here but only three parameters ($ g $, $ \alpha $ and $ \beta $) need to be fixed.
In the following study,  $ \alpha $ and $ \beta $  are set as free parameters and $ G_s^0 $ is constrained by low energy experimental data in the three-momentum cut-off regularization with $  G_s^0= 5.074 \times10^{-6} $ MeV. As $\alpha$ and $\beta$ are given, the coupling $g$ is derived and thus the other couplings. The other parameters  are current quark mass $ m = 5.5$ MeV and cut-off $\Lambda  = 631$ MeV.

\subsection{Solutions in the Nambu-Gorkov formalism  }

In dealing the gap equations for colored quarks we   apply the Nambu-Gorkov formalism \cite{Nambu:1960tm} with

\begin{equation}
S^{-1}(p)=\left(\begin{array}{l}
\not{p}+\hat \mu \gamma^0-\hat M\quad \Delta\gamma_{5}\tau_2\lambda_2\\
    -\Delta^*\gamma_{5}\tau_2\lambda_2\quad \not p-\hat \mu \gamma^0-\hat M
\end{array}
\right),
\end{equation}
where $\hat{M}=M_0+M_8\lambda_8$ and $\hat\mu=\tilde{\mu}+\tilde{\mu}_8\lambda_8$ are defined in the bispinor space with
\begin{equation}
  \Psi(x)= \dfrac{1}{\sqrt{2}}\left(  \begin{array}{c}
  	q(x)\\q^C(x)
  		\end{array}\right)
\end{equation}
And in the mean field approximation, the Lagrangian  with chemical potential  $\mu$ is given by $ \overline\Psi S^{-1}  \Psi -V $,
with
\begin{equation}
V=\dfrac{(M_0-m)^2}{4G_s^{(0)}}+\dfrac{M_8^2}{4G_s^{(8)}}+\dfrac{(\tilde{\mu}-\mu)^2}{4G_v^{(0)}}+\dfrac{\tilde{\mu}_8^2}{4G_v^{(8)}}+\dfrac{|\Delta|^2}{4H}.
\end{equation}
The constituent quark masses are defined as
\begin{equation}
M_0=m-2G_s^{(0)}\phi, \quad M_8=m-2G_s^{(8)}\phi_8
\end{equation}
and the effective chemical potentials are
\begin{equation}
\tilde{\mu}=\mu+2G_v^{(0)}n, \quad \tilde{\mu}_8=2G_v^{(8)}n_8.
\end{equation}
with $ n=\left<\bar q\gamma^0 q\right> $ and $
n_{8}=\left\langle\bar{q} \gamma^{0} \lambda_{8} q\right\rangle=\frac{2}{\sqrt{3}}\left(n_{r}-n_{b}\right)
$.

The diquark gap is given by the expectation value of  $ q^T i\gamma_5C\tau_2\lambda_A q $, with

\begin{equation}
\Delta=-2H\delta=-2H\left<q^T i\gamma_5C\tau_2\lambda_A q\right>.
\end{equation}
The presence of $\delta$ brakes the color-$SU(3) $, and only the red and green quarks paticipate in the condensate. The quark condensates of red and blue quarks are expected to be different.
 Introducing the colored quark mass and color chemical potential
  \begin{eqnarray}
 M_r=M_0+\dfrac{1}{\sqrt{3}}M_8,&&M_b=M_0-\dfrac{2}{\sqrt{3}}M_8,\\
 \phi_8=\dfrac{2}{\sqrt{3}} (\phi_r-\phi_b),&&   \tilde\mu_8=\dfrac{2}{\sqrt{3}} (\tilde\mu_r-\tilde\mu_b),
 \end{eqnarray}
we have
\begin{eqnarray} \begin{array}{l}
 M_r=m-\frac{2}{3}(6G_s^{(0)}+2G_s^{(8)})\phi_r-\frac{2}{3}(3G_s^{(0)}-2G_s^{(8)})\phi_b, \\
 M_b=m-\frac{2}{3}(6G_s^{(0)}-4G_s^{(8)})\phi_r-\frac{2}{3}(3G_s^{(0)}+4G_s^{(8)})\phi_b, \\
 \tilde{\mu}_r=\mu+\frac{2}{3}(6G_v^{(0)}+2G_v^{(8)})n_r+\frac{2}{3}(3G_v^{(0)}-2G_v^{(8)})n_b, \\
 \tilde{\mu}_b=\mu+\frac{2}{3}(6G_v^{(0)}-4G_v^{(8)})n_r+\frac{2}{3}(3G_v^{(0)}+4G_v^{(8)})n_b.
\end{array}
\end{eqnarray}
The thermodynamic potential by evaluating the trace and performing the Matsubara sum at nonzero temperature is given by

\begin{equation}
\begin{aligned}
\Omega(T, \mu)=&-4 \int \frac{\mathrm{d}^{3} p}{(2 \pi)^{3}}\left\{2\left(\frac{\omega_{-}+\omega_{+}}{2}+T \ln \left(1+\mathrm{e}^{-\omega_{-} / T}\right)\right.\right. \\
  + T&\left. \ln \left(1+\mathrm{e}^{-\omega_{+} / T}\right)\right)
  +\left(E_{p, b}+T \ln \left(1+\mathrm{e}^{-E_{-} / T}\right)\right.\\
   +T&\left.\left. \ln \left(1+\mathrm{e}^{-E_{+} / T}\right)\right)\right\}+V+\mathrm{const.}
\end{aligned}
\end{equation}
Here, $E_{p,b}=\sqrt{\vec{p}^2+M_b^2}$ and  the dispersion laws for colored quarks read as
\begin{eqnarray}
\omega_{\pm}=
	\sqrt{(\sqrt{\vec{p}^{2}+M_{r}^{2}}\pm\tilde{\mu}_{r})^{2}+|\Delta|^{2} },
 \end{eqnarray}
which produces gap of  $2\Delta$  for the coupled quark in the particle-hole excitation spectrum.

The stationary points of the thermodynamic potential give the expectation values of condensates which are

\begin{eqnarray}
\phi_{r}&=&-4 \int \frac{\mathrm{d}^{3} p}{(2 \pi)^{3}}\left\{\frac{M_{r} s-\tilde{\mu}_{r} t}{2 s \omega_{-}} \tanh \left(\frac{\omega_{-}}{2 T}\right)\right. \nonumber\\
 &&\quad~~~ +\left.\frac{M_{r} s+\tilde{\mu}_{r} t}{2 s \omega_{+}} \tanh \left(\frac{\omega_{+}}{2 T}\right)\right\} \\
\phi_{b}&=&-4 \int \frac{\mathrm{d}^{3} p}{(2 \pi)^{3}} \frac{M_{b}}{E_{p, b}}\left(1-n_{p, b}\left(T, \tilde{\mu}_{b}\right)\right.\nonumber\\
&&\quad~~~ -\left.\bar{n}_{p, b}\left(T, \tilde{\mu}_{b}\right)\right) \\
n_{r}&=&4 \int \frac{\mathrm{d}^{3} p}{(2 \pi)^{3}}\left\{\frac{\tilde{\mu}_{r}\left(s-\vec{p}^{2}\right)-M_{r} t}{2 s \omega_{-}} \tanh \left(\frac{\omega_{-}}{2 T}\right)\right.\nonumber\\
&&\quad~~~  +\left.\frac{\tilde{\mu}_{r}\left(s+\vec{p}^{2}\right)+M_{r} t}{2 s \omega_{+}} \tanh \left(\frac{\omega_{+}}{2 T}\right)\right\} \\
n_{b}&=&4 \int \frac{\mathrm{d}^{3} p}{(2 \pi)^{3}}\left(n_{p, b}\left(T, \tilde{\mu}_{b}\right)-\bar{n}_{p, b}\left(T, \tilde{\mu}_{b}\right)\right) \\
\delta&=&-8 \int \frac{\mathrm{d}^{3} p}{(2 \pi)^{3}}\left\{\frac{\Delta s+\Delta_{0} t}{2 s \omega_{-}} \tanh \left(\frac{\omega_{-}}{2 T}\right)\right.\nonumber\\
&&\quad~~~
+\left.\frac{\Delta s-\Delta_{0} t}{2 s \omega_{+}} \tanh \left(\frac{\omega_{+}}{2 T}\right)\right\}
 \end{eqnarray}
The blue quark is unpaired, and here  $n_{p,b}$ and  $\bar n_{p,b}$ are the usual Fermi occupation functions for the blue quarks and antiquarks.
With the parameters given before, these self-consistent equations give the quark mass about $ 267 $ MeV  at zero temperature and zero chemical potential.


\section{Results} \label{sec.result}

\subsection{The chiral phase transition }
\begin{figure}[h]
	\begin{center}
		\includegraphics[width=0.45\textwidth]{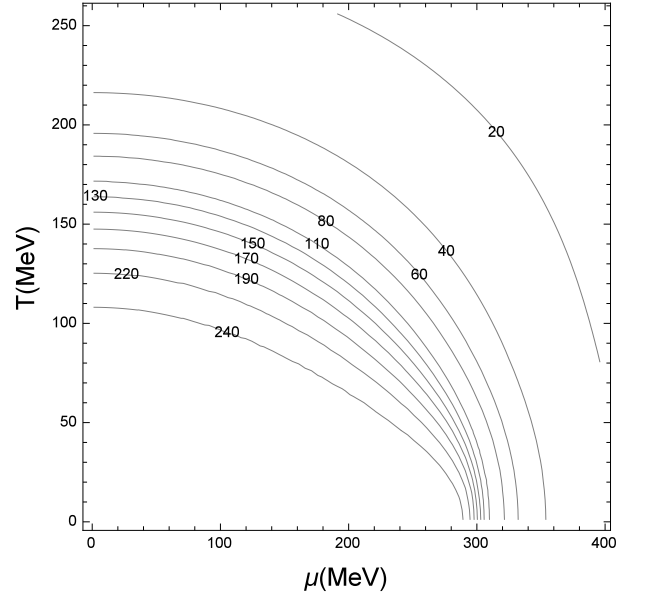}	\caption{Contour plot of quark mass as function of chemical potential and temperature of equal weights when $\alpha =\beta=1/3$.}\label{fig.cont}
	\end{center}
\end{figure}

With the given parameters under three-momentum cut-off regularization, the chiral phase transition is first order under low temperature and crossover under high temperature. The quark mass as function of chemical potential and temperature is presented in Fig. \ref{fig.cont} with equal weight for the three parts of interaction.

\begin{figure}[htp]
	\begin{center}
		\includegraphics[width=0.45\textwidth]{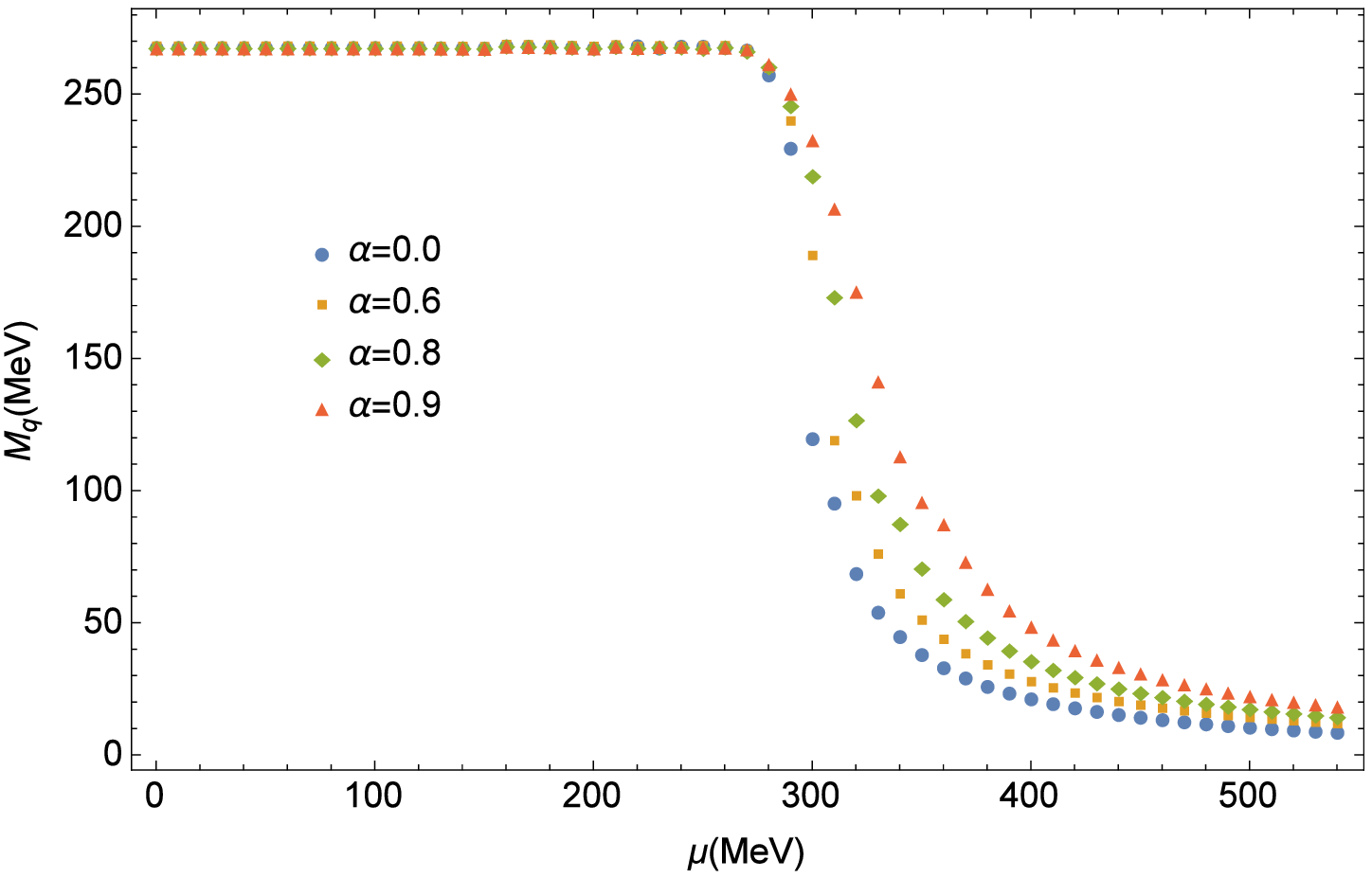}\\	\includegraphics[width=0.45\textwidth]{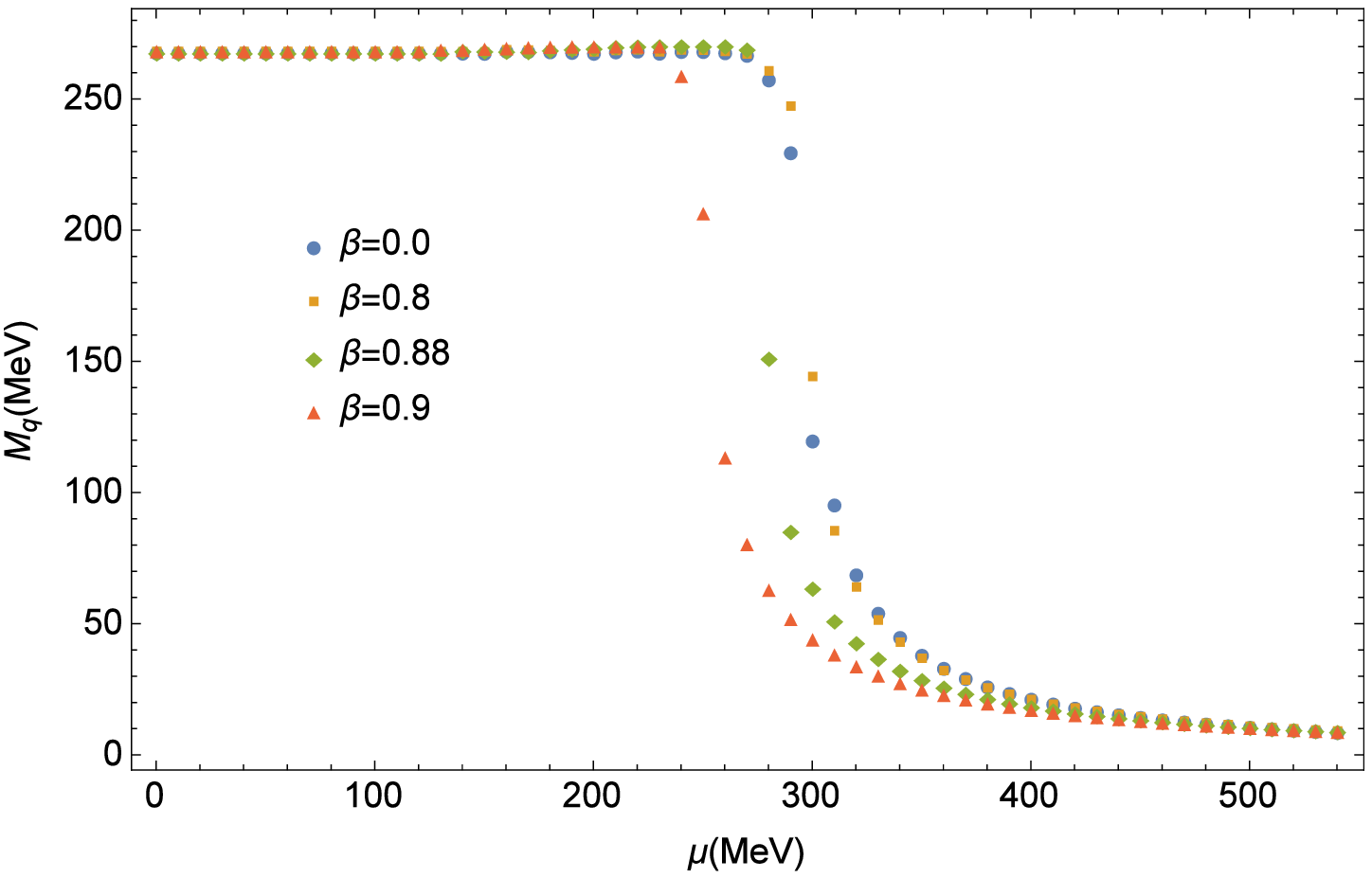}		
		\caption{Quark mass as functions of weight parameters $\alpha$ $(\beta=0)$ and $\beta$ $(\alpha=0)$, respectively. }\label{fig.mq}
	\end{center}
\end{figure}

The first order transition will turn to   crossover as the
the vector channel is more involved with large $\alpha$, as shown in the upper panel of Fig. \ref{fig.mq}. More discussions on the   effect of the vector channel can be found in previous works \cite{Wang:2019uwl,Li:2018ltg, Wang:2019jze}.

The first order phase transition still remains as we  increase the weight of diquark interaction channel. As $\beta$  increases the pseudo-critical chemical potential decreases. In the lower panel of Fig. \ref{fig.mq}, the obvious change occurs when $\beta$ is larger than 0.8.  When $\alpha=0$ and $\beta=0.5$ the ratio  of coupling strength of the scalar quark-antiquark channel and diquark channel is $G_s^{(0)}:H=8:1$ and thus the diquark channel heavily suppressed. So to manifest the colored diquark effect on the phase transition, the weight $\beta$ must be larger than that.

\subsection{Mass and quark number difference}
The Fierz-transformed Lagrangian $\mathscr{L}_{ex}$ has included both the scalar channel and vector channel, the increase of $\alpha$ will also increase the scalar interaction.
We can study the competition between scalar channel and vector channel by fixing $\beta$. In order to reflect the competition between vector channel and colored diquark  interaction channel more clearly, we fix the weight of direct interaction term $\mathscr{L}_{int}$  and study the results under different $\alpha$ and $\beta$.

The 2SC model does not guarantee that the color neutrality is  fulfilled. The red and blue quark have different effective masses and number densities as the gap appears to be nonzero.

 \begin{figure}[h]
	\begin{center}
		\includegraphics[width=0.45\textwidth]{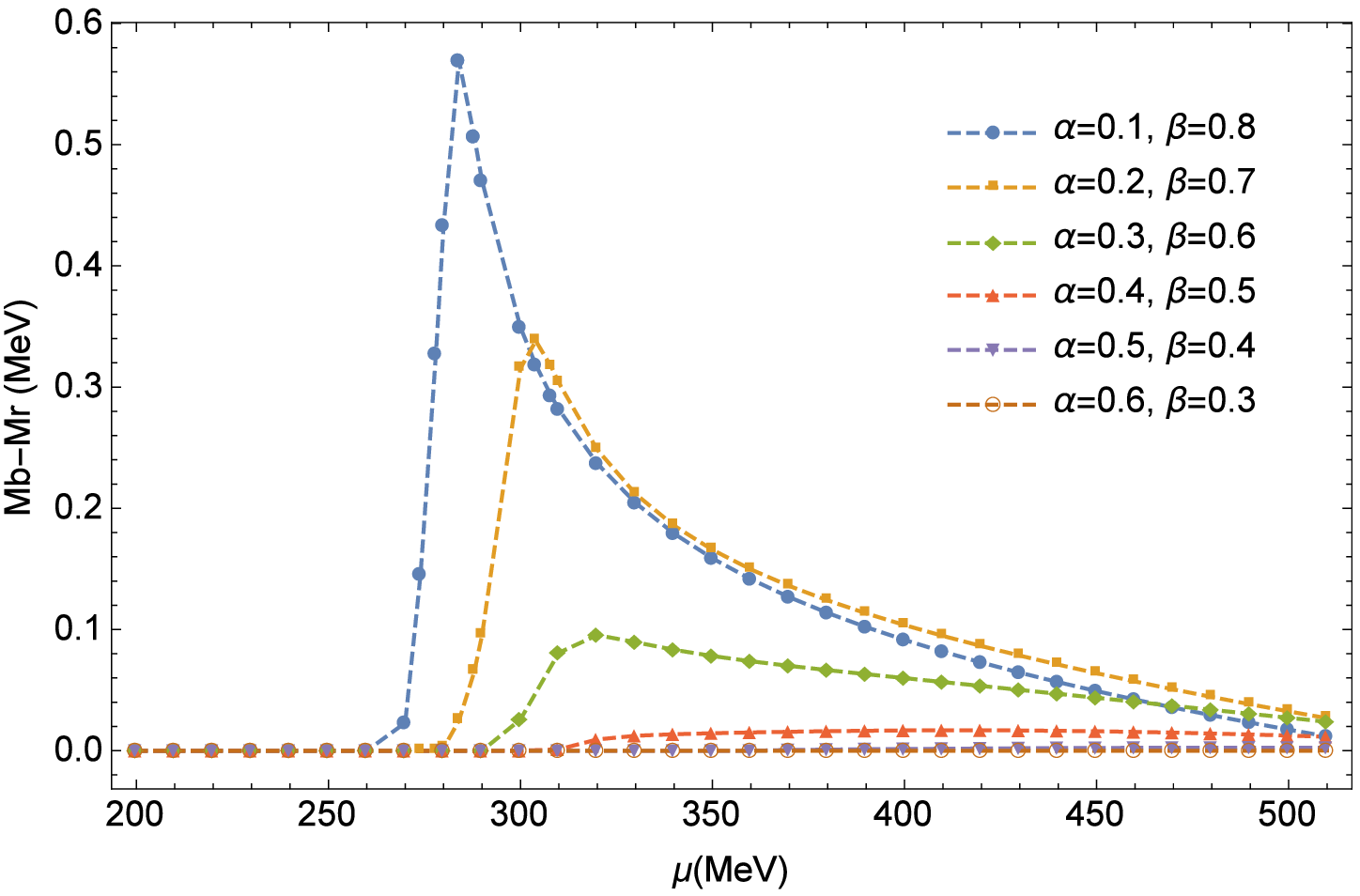}\\	\includegraphics[width=0.45\textwidth]{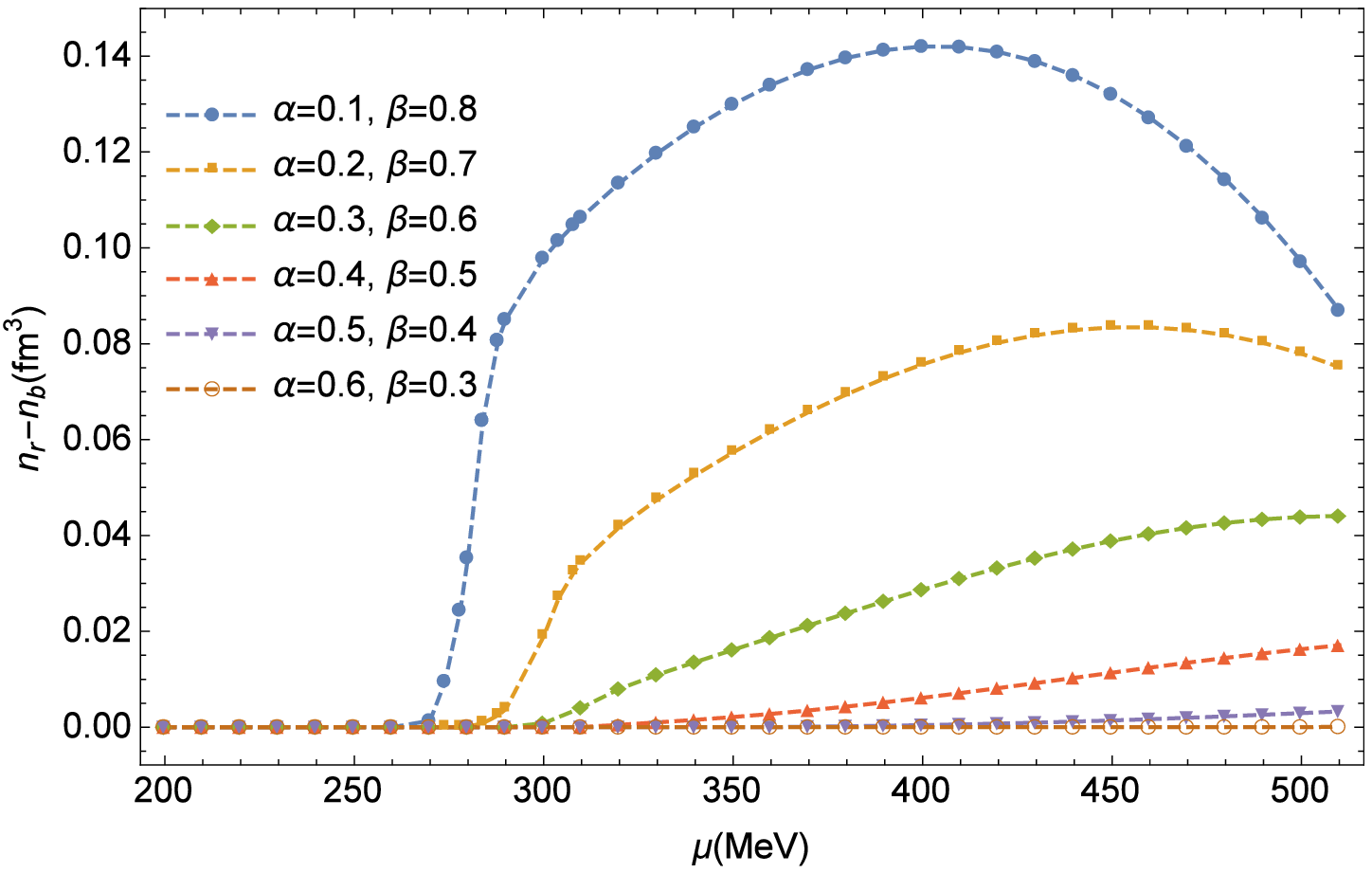}		
		\caption{ The mass and quark number density at different $\alpha$ and $\beta$ as function of chemical potential.  }\label{fig.d}
	\end{center}
\end{figure}

At low chemical potential the mass and quark number density are equal for red and blue quark but  increase  rapidly  when the chemical potential is larger than some specific values, as shown in Fig. \ref{fig.d}.  The mass difference is less than $ 1 $ MeV. Since the chiral symmetry is partly restored at high chemical potential,   the mass difference  drops to almost zero.
The difference of number density increases as $\beta$ increases.  We can also see in Fig. \ref{fig.d} that  the difference of number density decreases at high chemical potential for $\beta=0.8$.
It   may indicates that the rate of formation of quark pairs should be reduced at high chemical potential. At more large chemical potential the model may become sensitive  to the cut-off and finally  it reaches the limits of the model.

\subsection{  Gap and specific heat }

\begin{figure}[h]
	\begin{center}
		\includegraphics[width=0.45\textwidth]{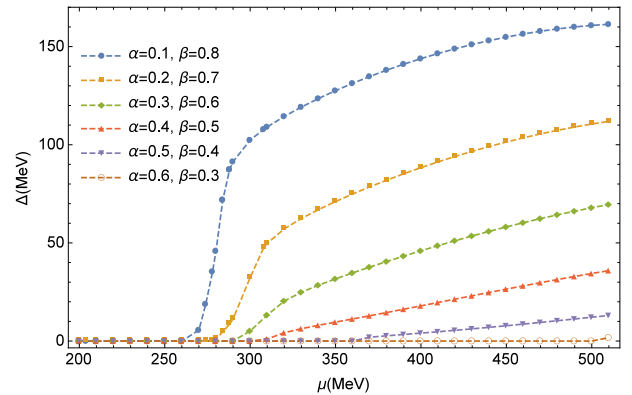}	
		\caption{Diquark condensate as function of chemical potential at zero temperature.}\label{fig.deltah}
	\end{center}
\end{figure}

\begin{figure}[h]
	\begin{center}
		\includegraphics[width=0.45\textwidth]{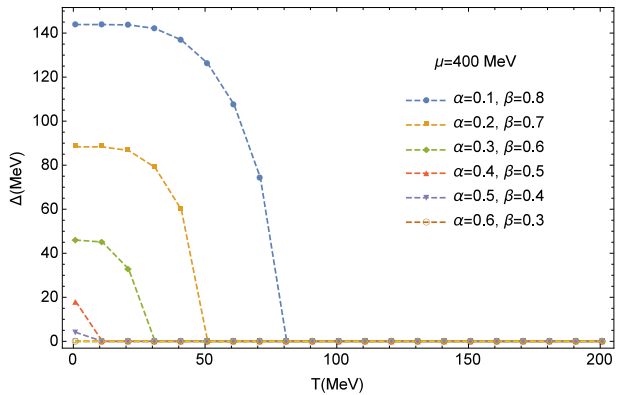}
		\caption{Diquark condensate as function of temperature at  $\mu=400$ MeV.}\label{fig.deltat}
	\end{center}
\end{figure}

A direct manifestation of color super conduct of super fluid is the  appearance of a non-negligible gap in the dispersion law.  The gap $\Delta$ varies with temperature and chemical potential. The calculated results are presented in Fig. \ref{fig.deltah} and  \ref{fig.deltat} . The gap increases with chemical potential and $\beta$,  and  only for relative large $\beta$  the interactions give rise to a non-negligible gap. Since  the Lagrangian is dominated by the Fierz-transformed diquark part, the gap $\Delta$ can be larger than 100 MeV. Critical chemical potential exists in the Fig. \ref{fig.deltah} and before critical point the gap $\Delta$ is zero.

\begin{figure}[h]
 	\begin{center}
 		 		\includegraphics[width=0.45\textwidth]{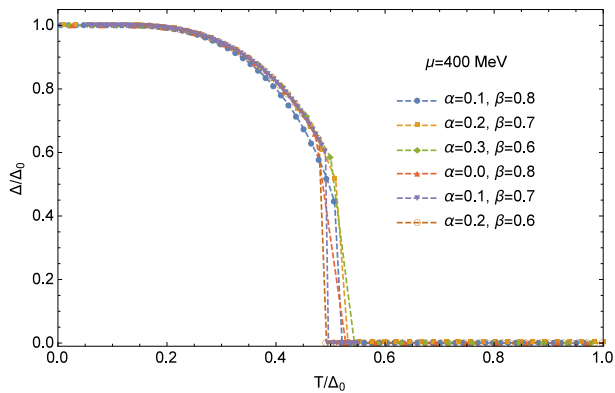}
 		\caption{Normalized condensates as function of temperature. $\Delta_{0}$ is the gap at zero temperature. }\label{fig.deltatc}
 	\end{center}
 \end{figure}

The diquark could be  de-paired as temperature increases just the same as in the BSC theory for electrons. The diquark condensate vanishes and the gap decreases to zero as temperature increases.
There is a well-known BCS relation that
\begin{equation} \label{tcdelta}
T_c\simeq 0.57 \Delta_{0},
\end{equation}
where $\Delta_{0}$ is the gap at zero temperature and $ T_c $ is the temperature where gap begin to be zero \cite{Buballa:2001gj}. We plot the $\Delta/\Delta_0\sim T$ relation calculated at $\mu= 400$ MeV in Fig. \ref{fig.deltatc}. At various parameter set, the critical temperature =  $ T_c $  is  located between 0.5 and 0.6.

\begin{figure}[h]
	\begin{center}
		\includegraphics[width=0.45\textwidth]{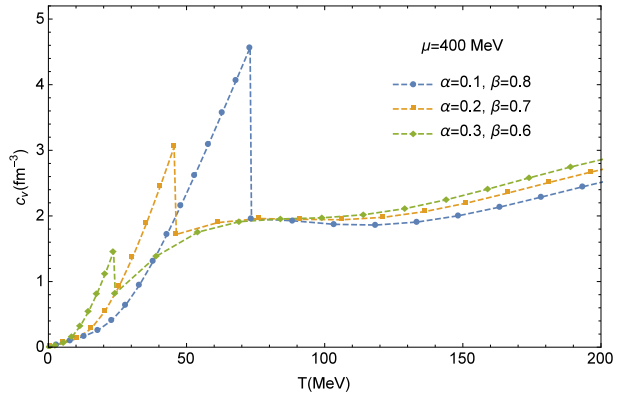}\\	
		\caption{Specific heat as function of temperature at fixed chemical potential.}\label{fig.sp}
	\end{center}
\end{figure}

Furthermore, we calculate the specific heat as function of temperature. At fixed chemical potential, the specific heat is given by

\begin{equation}
c_v=-T\dfrac{\partial^2\Omega}{\partial T^2}.
\end{equation}
We show the results in Fig. \ref{fig.sp}.  The specific heat increases from zero at low temperature. When the temperature continues to increase  the specific heat has a jump-down at large chemical potential and $\beta$. The jump-down occurs at the critical temperature. At fixed chemical potential,  the critical temperature increases with $\beta$.

\section{Summary}

 Without performing mean field approximation, the original Lagrangian and its Fierz transformation are numerically identical. The Fierz-transformed Lagrangian contains a variety of interaction channel.  But at the mean field level, only some specific terms remain. We take the idea that all these equal interaction should be included in one combined Lagrangian with different weights.

 In this paper, starting from the standard NJL Lagrangian with only scalar and pseudo scalar interaction other than from the color current interactions, we have firstly composed Lagrangian with its Fierz transformation.  The five coupling constants are replaced by two weighting parameters.  The Fierz-transformed Lagrangian contains both quark-antiquark interaction included the   vector channel interaction and quark-quark interaction.   In the quark-quark interaction we care about the scalar and pseudo-scalar channel, but only  scalar channel remain after the fierz transformation.

 The competition of the two transformed Lagrangian are studied.
 We have calculated the chiral phase transition. It shows that the vector channel and diquark channel interaction have opposite effect on the chiral transition as chemical potential increase. The vector channel interaction postpones the transition while   the diquark interaction advances the chiral phase transition as the chemical potential increases.

 We have also calculated the  colored quark mass and number density difference, the superconductor gap and the specific heat.  The mass difference of colored quark can be negligible.
But critical chemical potential exits where the  quark mass and number density difference begin to increase from zero and the gap $\Delta$ also increases. We have calculated the gap as function of temperature as specific  chemical potential for different weight parameters.  It is found that although the critical temperature and gaps are varied but the normalized gap as function of temperature are almost coincides. The ratios of   critical temperature and the gap at zero temperature are located between 0.5 and 0.6 which are in consistent with   the BCS relation with the ratio of  0.57.

In the end we confirm the phase transition to superconductor phase by the calculation of specific heat.


\acknowledgments
This work is supported by the Cultivating Plan of Characteristic Direction of Science (2020SCUNL209).

\end{document}